\documentclass[12pt]{article}
\usepackage{xurl}
\usepackage[margin=1in]{geometry}
\usepackage{graphicx}
\usepackage{amsmath}
\usepackage{amsfonts}
\usepackage{amsthm}
\usepackage{hyperref}

\usepackage{bm}
\usepackage[usenames,dvipsnames,svgnames,table]{xcolor}

\begin{document}

\title{Weights and Methodology Brief for the COVID-19 Symptom Survey by University of Maryland and Carnegie Mellon University, in Partnership with Facebook}
\author{Neta Barkay, Curtiss Cobb, Roee Eilat, Tal Galili, Daniel Haimovich, \\ Sarah LaRocca, Katherine Morris, Tal Sarig}
\date{
    \href{mailto:COVID19symptomsurvey@fb.com}{Facebook Research} \\
    \bigskip
    \smallskip
    September 4, 2020, Version 3
}

\maketitle

\section{Introduction}

The Facebook company is partnering with academic institutions to support COVID-19 research and to help inform public health decisions. Currently, we are inviting Facebook app users in the United States to take a survey collected by faculty at Carnegie Mellon University (CMU) Delphi Research Center, and we are inviting Facebook app users in more than 200 countries or territories globally to take a survey collected by faculty at the University of Maryland (UMD) Joint Program in Survey Methodology. As part of this initiative, we are applying best practices from survey statistics to design and execute two components:

\begin{enumerate}
\item Sampling Design: deciding who to invite to participate in the survey each day.
\item Weighting Methodology: providing a weight per user so that respondents better represent the target population as a whole.
\end{enumerate}

We and our partners designed this initiative with privacy in mind from the start. The survey and its privacy practices are reviewed by the Institutional Review Boards of both UMD and CMU. Facebook does not receive any survey responses to weight the data. Instead, UMD and CMU send Facebook the list of Random ID numbers for the users who complete the survey each day. We then use internal Facebook data covered by our Data Policy in conjunction with publicly available population benchmark data to calculate a single weight for each user in the survey sample., We then provide these weights only to researchers with an approved Data Use Agreement.\footnote{https://www.facebook.com/policy.php} We and our partners describe the surveys, including the privacy-preserving processes we use to weight the data, in a special issue of Survey Research Methods (Kreuter et al. 2020). 

Using the total survey error framework (Groves and Lyberg 2010 \cite{groves_2010}), our goal when calculating the weights is to minimize errors of representation, including coverage, random sampling and non-response errors. We achieve this through generating weights in two stages for the US CMU and global UMD surveys. First, we adjust for non-response error using Inverse Propensity Score Weighting (IPSW) to make the sample more representative of the sampling frame of Facebook app users. Second, we adjust for coverage error using post-stratification with weights from the first stage as inputs. Intuitively, the final weights can be understood as the number of adults in the general population who are represented by a respondent in the sample for that day.\footnote{In the countries or territories where we include administrative regions in the post-stratification step (e.g., US states), the final weights can be used to calculate representative statistics at either the administrative region or country-level. There are some countries or territories where we do not include administrative regions in the post-stratification step, and in these places, the final weights can be used to calculate representative statistics at only the country-level. Please see ``Representing the general adult population" for more information.} A respondent who belongs to a demographic group that has a high likelihood of responding to the survey may get a weight of 100 while someone who belongs to a group that is less likely to respond may get a weight of 500.\footnote{Early versions of the US weights were divided by the overall adult population size of the US to obtain very small values. This was a temporary technical decision that was reversed. If the user still has these earlier versions of the weights, we encourage use of the new weight values or undoing this technical decision by multiplying the earlier version of the weights by 249,194,000.}

The weights are available for the CMU US survey and, separately, for 114 other countries or territories in the UMD global survey. The set of non-US entities for which we provide weights was determined by our ability to generate high quality results as well as other considerations. Aggregate weighted estimates are publicly available through UMD and CMU.\footnote{\href{https://covidmap.umd.edu/api.html}{https://covidmap.umd.edu/api.html} \cite{fan_2020}; \newline \href{https://cmu-delphi.github.io/delphi-epidata/api/covidcast.html}{https://cmu-delphi.github.io/delphi-epidata/api/covidcast.html} \cite{farrow_2015}} Academic and nonprofit researchers may request access to non-aggregated survey data in addition to the raw survey weights for their research. Once an initial request is approved by both Facebook and either UMD or CMU, the researcher’s institution must then sign Data Use Agreements before data access will be provided by UMD or CMU. More information can be found on the Facebook Data for Good website.\footnote{https://dataforgood.fb.com/docs/covid-19-symptom-survey-request-for-data-access/}

Below we provide a more technical overview of the methodology, our choices, and provide guidelines for using the survey weights.

\section{Sampling Design}

\paragraph{Sampling frame.} The sampling frame is the Facebook Active User Base (FAUB) ages 18+, which includes users living within 200+ countries or territories. The sampling frame is restricted to people who use Facebook in one of the supported locales.\footnote{A locale is composed of a base language in combination with a geography or dialect. For example, English in the United States is ``en\_US" whereas English in the United Kingdom is ``en\_GB.'' The CMU US survey was launched on April 22, 2020 in English only, including en\_US and en\_GB; it was made available in the additional 6 locales on May 22, 2020: Spanish - Spain (es\_ES), Spanish - Latin America (es\_LA), French - France (fr\_FR), Brazilian Portuguese (pt\_BR), Vietnamese - Vietnam (vi\_VN), and simplified Chinese (zh\_CN).} At this time, the CMU US survey is available in 8 locales and the UMD international survey is available in 55 locales.. While this affects coverage, it ensures that the entire communication from invitation to instrument is in the user's language. The proportion of users who are ineligible due to this restriction is less than 5\% globally, though this varies by country or territory. While the sampling frame does not cover the population in all countries or territories where the survey is fielded, it covers a large proportion of the global population.\footnote{The Facebook monthly active users figure is reported in the company’s quarterly earnings report found on https://investor.fb.com. As of March 31, 2020, there were 2.6 billion monthly active users globally, including 186 million monthly active users in the US and Canada.}

\paragraph{Sampling method.} The surveys are daily repeated cross-sections, with similar user characteristics across days. The Facebook app invites a new sample of adult users to take the survey each day. Sampled users see an invitation at the top of their Facebook News Feed to an optional, off-Facebook survey. In order to provide geographic coverage, we use stratified random sampling using administrative boundaries within countries or territories. And, in order to reduce survey fatigue over the course of field collection, we employ differential sampling probabilities across these administrative boundaries. Sampled users may be invited to take the survey again in either a few weeks or months, depending on the population density of their area. In low density regions, eligible Facebook users are sampled into the survey once a month. In high density regions, eligible Facebook users are sampled every two to six months. The survey responses of sampled users who participate more than once cannot be linked longitudinally.  

\section{Weighting Methodology}

\paragraph{Defining the weighted sample.} We provide weights for two sets of sample respondents separately for both the CMU US and UMD global surveys. First, we provide weights for respondents who answered the questions needed to calculate the aggregate estimates of COVID-like Illness (CLI) reported in the CMU and UMD APIs. Second, we provide weights for a larger set of respondents who answered a minimum of two questions in the surveys.

\paragraph{Considerations in choosing the methodology.} Recognizing that multiple researchers will be conducting analyses without our direct involvement, we are taking a conservative approach and prioritize simplicity and well-established principles over sophistication. Our goal is for the weights to be used in a straightforward manner to improve the accuracy of the results of our academic partners without introducing any difficulties. As noted below in the section, ``Using the Weights," we expect that some researchers may want to implement further bias-correction to address their specific use cases.

\paragraph{Representing the FAUB.} We apply IPSW to represent adult FAUB using nonresponse weights. As covariates, we use existing attributes which are generally available for Facebook users. We choose IPSW as a well-established approach which is both simple and allows correcting for many covariates simultaneously. We transform continuous variables into multiple buckets to ensure that we better match their full distribution rather than the mean only. We also apply regularization to our model to minimize the variance of the weights (both within users and between users), and apply a standard trimming procedure for the weights. This method of bias correction results in a good representation of the FAUB which, as already noted, constitutes a sizable proportion of the global population.

\paragraph{Covariates to model non-response within the FAUB sampling frame.} The covariates used to model non-response do not come from individual survey responses, which Facebook does not collect or receive, but rather come from internal Facebook data. We include self-reported age and gender as well as other attributes which we have found in the past to correlate with survey outcomes. We believe that these attributes correlate to other demographic attributes which are not available to us, implicitly improving overall demographic representation without attempting to infer them. We also include some geographical variables to improve geographic representation. The choice of covariates is guided by the discussion in Little and Vartivarian (2005) \cite{little_2005}.

\paragraph{Representing the general adult population.} To improve representation of the entire adult population in a country or territory, including people not covered by the FAUB sampling frame, we apply a second step of Post-Stratification (PS) using publicly available benchmarks and the IPSW output weights as input weights to this stage. For the US, we post-stratify over state, age, and gender using benchmarks obtained from the Current Population Survey 2018 March Supplement \cite{cps_2020}.\footnote{Age buckets of 18-24, 25-44, 45-64, 65+.} For other countries or territories, we rake over age and gender using benchmarks obtained from the United Nations (UN) Population Division 2019 World Population Projections, and first administrative level regions (commonly referred to as ``subnational regions") using benchmarks constructed from publicly available population density maps \cite{unpd_2020}. \footnote{The first administrative level population benchmarks are constructed from two population density map sources: The first is the High Resolution Settlement Layer (HRSL) from a partnership between Facebook and the Center for International Earth Science Information Network (CIESN) at Columbia University, which is distributed through the Facebook Data for Good program. The second is the Gridded Population of the World, Version 4 (GPWv4) from CIESIN \cite{ciesin_2018}. These data have been used by non-governmental organizations for various public health projects, including by the Red Cross (Facebook 2019 \cite{facebook_ai_2019}).} Though, there are several countries or territories where we only post-stratify by age and gender and do not rake over administrative level geographies.\footnote{In the following list of countries we do not correct for region because of small sample size or high imbalance of respondents between regions: AF, AL, AM, AO, AZ, BA, BF, BJ, BY, CD, CI, CM, DO, ET, GH, GN, HK, HT, KE, KG, KH, KW, KZ, LA, LB, LK, LY, MD, MG, ML, MM, MR, MZ, NI, NP, OM, PS, QA, RO, SD, SN, TZ, UZ, YE. Instead, we post stratify over age and gender only.} This ensures good representation over these important variables\footnote{In the UMD global survey, there are some days of data from some countries or territories over the course of field collection where there are no respondents in a given region or age-gender bucket used in raking. In these instances, the regions or age-gender buckets without respondents are omitted and the weights are scaled to the remaining population size. This means, for example, that if a given region does not have respondents, it will be omitted from the raking procedure, it will not be represented in the country-level aggregation, and the country-level weights will not sum up to the country or territory's adult population size.} Finally, weights are trimmed from the bottom and the top for each country or territory.  When a weight is smaller than the mean weight of that country or territory divided by 30 or larger than 10 times the mean weight, that weight is truncated to these floors/ceilings and the sum of weights are rescaled to the population size from the UN Population Projections. To the extent that other demographic variables are available to data users through the survey responses, we encourage them to check representation of these variables against other population benchmarks and consider applying additional weighting steps to address any differences.

\section{Using the Weights}

Survey weights are fairly straightforward to use and many types of analyses can be easily adapted to work with a weighted sample. We provide guidance for the three most common estimators. To the best of our knowledge, these estimators will fully address the needs in this initiative. 

For each quantity of interest, we provide an estimator as well as a variance estimator that can be used to account for the random sampling error. This can be done by constructing approximate confidence intervals by taking a margin of $\alpha\hat V$ with a typical $\alpha$ value being $1.96$. To produce conservative estimates which may be more suitable for visualizing in a map, a heuristic such as using the lower/upper confidence bound may be used. We refer to the discussion in Gelman and Price (1999) \cite{gelman_1999}.

Let $w$ be the weights provided by us, $T$ be the target population,\footnote{In countries or territories where we include administrative regions in the post-stratification or raking step, the target population might be either a region or the parent country or territory. In countries or territories where we do not include administrative regions in the raking step, the target population is the country or territory. See the footnote above for more details.} $S$ the sample, and $y,z$ be different outcomes of interest. We quote these results from Särndal, Swensson, and Wretman 2003 (p. 178) \cite{sarndal_2003}.

\paragraph{Population mean.} Consider the population mean $\bar y = \frac{1}{\left| T \right|} \sum_{i \in T} y_i$. To estimate it use 

$$ \hat y = \frac{\sum_{j \in S} w_j y_j}{ \sum_{j \in S} w_j}$$

If an estimator for the variance of the population mean is needed, use 

$$ \hat V = \frac{\sum_{j \in S} w_j^2 (y_j - \hat y)^2}{ ( \sum_{j \in S} w_j )^2} $$

\paragraph{Population total.} To estimate the total $ t = \sum_{i \in T} y_i$ use 

$$ \hat t = \sum_{j \in S} w_j y_j$$

and 

$$ \hat V = \sum_{j \in S} w_j^2 (y_j - \hat y)^2$$

\paragraph{Population ratio.} To estimate the ratio $ r = \sum_{i \in T} y_i / \sum_{i \in T} z_i$ use 

$$ \hat r = \frac{\sum_{j \in S} w_j y_j}{\sum_{j \in S} w_j z_j } $$ 

and 

$$ \hat V = \frac{\sum_{j \in S} w_j^2 (y_j - \hat r z_j)^2}{(\sum_{j \in S}{w_j z_j})^2} $$

We expect the population ratio estimator to be used most often, as when estimating quantities such as the fraction of COVID-like illness ($y_j = 1$ if the respondent has qualifying symptoms) in a geographical region ($z_j = 1$ if the respondent is in the target region). We also suggest that parametric approaches or small area estimation may be appropriate in regions where too few responses are available.

\bibliographystyle{plain}
\bibliography{references}

\end{document}